\documentclass[%
 reprint,
superscriptaddress,
 amsmath,amssymb,
 aps,
pra,
]{revtex4-1}

\usepackage{latexsym,amsmath,amssymb}
\usepackage{braket} 
\newcommand{\ketbra}[2]{\ket{#1}\hspace{-2.5pt}\bra{#2}}
\usepackage{ftnxtra}
\usepackage{fnpos}
\usepackage{tikz}
\usepackage{color}
\usepackage{footnote}

\usepackage{amsthm} 

\newtheorem{result}{Result}

\usepackage[T1]{fontenc} 

\usepackage{graphicx}
\usepackage{epstopdf}
\newcommand{\Tr}{\mathrm{Tr}}
\newcommand{\swap}{SWAP }
\newcommand{\ba}{\begin{eqnarray}}
\newcommand{\ea}{\end{eqnarray}}
\newcommand{\ban}{\begin{eqnarray*}}
\newcommand{\ean}{\end{eqnarray*}}

\begin{document}
\title{Self-testing using only marginal information}
\date{\today}
\author{Xinhui Li}
 \affiliation{State Key Laboratory of Networking and Switching Technology, 
Beijing University of Posts and Telecommunications, Beijing, China 100876}
 \affiliation{Centre for Quantum Technologies, National University of Singapore, 3 Science Drive 2, Singapore 117543}

\author{Yu Cai}
 \email{caiyu01@gmail.com}
 \affiliation{Centre for Quantum Technologies, National University of Singapore, 3 Science Drive 2, Singapore 117543}

\author{Yunguang Han}
 \affiliation{Centre for Quantum Technologies, National University of Singapore, 3 Science Drive 2, Singapore 117543}

\author{Qiaoyan Wen}
 \affiliation{State Key Laboratory of Networking and Switching Technology, 
Beijing University of Posts and Telecommunications, Beijing, China 100876}

\author{Valerio Scarani}
 \affiliation{Centre for Quantum Technologies, National University of Singapore, 3 Science Drive 2, Singapore 117543}
 \affiliation{Department of Physics, National University of Singapore, 2 Science Drive 3, 117542, Singapore}

\begin{abstract}
The partial states of a multipartite quantum state may carry a lot of information: in some cases, they determine the global state uniquely. This result is known for tomographic information, that is for fully characterized measurements. We extend it to the device-independent framework by exhibiting sets of two-party correlations that self-test pure three-qubit states.


\end{abstract}

\maketitle


\section{Introduction}
\label{sec:introduction}
One of the most basic tasks in quantum information processing is to describe the state in the experiment, also known as state tomography. In the usual tomographic scenario, one would have access to a set of well characterized measurement devices. By repeating the experiment, expectation values of an informationally complete set of measurements allows us to reconstruct the density operator that describes the quantum state. 
In a multipartite scenario, it is sometimes possible to reconstruct the state with only marginal statistics~\cite{Linden1,Linden2,xin2017quantum,walck2008only,vertesi2014certifying}. For example, the three-qubit $W$ state
\ba\ket{W} = \frac{1}{\sqrt3}(\ket{001} + \ket{010} + \ket{100})\label{Wstate}\ea is the only state, pure or mixed, with partial states $\rho_{AB} = \rho_{AC} = \Tr_{B} \ketbra{W}{W}$. Thus, given those partial states, the global state can be inferred.

Remarkably, even when the devices are completely uncharacterized, an analog of tomography may be possible in the presence of Bell nonlocality: \textit{device-independent (DI) self-testing} \cite{mayers2003self}. Indeed, some nonlocal statistics identify one pure state and one set of measurements, up to local isometries; and if the observed statistics deviate from the ideal ones, one can estimate how far the actual state and measurements are from the ideal ones, a property known as \emph{robustness}. Among several recent results, it was proved that every pure bipartite entangled state can be self-tested \cite{coladangelo2017all}. For multipartite pure states, only examples are known: the self-testing of the $W$ state was first reported in Refs.~\cite{Wu,Pal}; for an updated list, see Refs.~\cite{fadel2017self,vsupic2017simple} and references therein. All these examples exploit correlations involving all the parties. In this paper, we show that, as it happens for tomography, it is possible to self-test some multipartite states using only marginal information.

The possibility of obtaining relevant information from partial correlations is paramount in many-body physics, where usually even three-body 
correlators are hard to measure. In this context, Bell inequalities that use only few-body correlators have recently attracted a lot of attention~\cite{Tura,Schmied}. At a more fundamental level: the proof that, if one were to simulate quantum entanglement with communication, this communication should travel at infinite speed, also relies on finding Bell inequalities using only marginal information~\cite{Bancal1,Barnea1}. Our work makes DI self-testing relevant for such studies.

Before presenting our four specific results of self-testing using only marginal information, we review the so-called \swap method, developed in~\cite{Bancal,Yang1} and based on a semidefinite optimization, that we are going to use.

\section{Tool: the \swap method}
\label{sec:swap}

Let us consider a Bell-type experiment involving a number of noncommunicating parties---for definiteness and for the sake of our specific results, we stay with three parties. Each has access to a black box with inputs $x,y,z \in \left\{0,1,...,M-1\right\}$ and outputs $a,b,c \in \left\{0,1,...,m-1\right\}$. Assuming quantum mechanics, one could model these boxes with an underlying state $\ket{\Psi}_{A,B,C}$ and measurement projectors $\left\{M_x^a\right\}_{x,a}$, $\left\{M_y^b\right\}_{y,b}$, and $\left\{M_z^c\right\}_{z,c}$, which commute for different parties. The state can be taken pure and the measurement projective without loss of generality, because the dimension of the Hilbert space is not fixed and the possible purification and/or auxiliary systems can be given to any of the parties. After sufficiently many repetitions of the experiment one can estimate the joint conditional statistics, also known as \emph{the behavior}, $p(a,b,c| x,y,z) = \bra{\Psi} M_x^aM_y^bM_z^c \ket{\Psi}$.

A device-independent certification is one that extracts nontrivial information on the state and the measurements from the behavior, without assumptions on the underlying degrees of freedom. In the case of device-independent self-testing, one wants to quantify the closeness of the unknown state used in the experiment $\ket{\Psi}$ to a desired target state $\ket{\psi}$. The idea of the \swap method is to ``swap'' out the essential information on to auxiliary systems with the same dimensionality as the local systems of the target state (here, we assume qubits). Specifically, the virtual protocol that one considers is the following.
\begin{enumerate}
    \item Distribute the state $\ket{\Psi}_{ABC}$, which produces the observed behavior, to Alice, Bob and Charlie.
    \item Alice, Bob and Charlie also have access to an auxiliary qubit, initialized in the state $\ket{000}_{A'B'C'}$.
    \item Alice, Bob and Charlie apply a local unitary, $U=U_{AA'}\otimes U_{BB'}\otimes U_{CC'}$, between their part of the unknown system and their auxiliary qubit.
\end{enumerate}
The closeness of the unknown resource to the target state can be then captured by the fidelity
$$f = \bra{\psi} \rho_{swap} \ket{\psi}\,,\label{eq:fidelity}$$
where
$$\rho_{swap}=\Tr_{ABC}(U\ketbra{\Psi}{\Psi}_{ABC}\otimes|000\rangle\langle 000|_{A'B'C'}U^\dagger).$$

The unitaries $U_{AA'}$ must be formally constructed with the unknown measurement operators. Then, $f$ becomes a linear combination of two types of terms: some that enter the observed behavior, and some non-observable correlations which involve different measurements on the same party, for example, $\bra{\Psi}M^a_{x}M^{a'}_{x'}\ket{\Psi}$ with $x\neq x'$, which are left as variables. Finally, with the aid of the Navascu\'es-Pironio-Ac\'in (NPA) hierarchy characterization of the quantum behaviors~\cite{navascues2008convergent}, a lower bound on $f$ can be computed as a semidefinite program (SDP):
\begin{align*}
    \min\quad &f=\bra{\psi}\rho _{swap}\ket{\psi} \\
    \text{s.t.}\quad &\Gamma \geq 0, \\
    &\Tr(\alpha_i\Gamma)=\delta_i, \quad i=1,2,...K
\end{align*}
where $\Gamma$ is the moment matrix of a certain level and matrices $\alpha_i$ and real number $\delta_i$ specifies the observed behavior. In self-testing by marginals, \textit{only marginal information on the behavior is specified in the constraints}: all terms involving three or more measurements, including observable ones, are left as SDP variables.

\section{Four results}
\begin{result}
The $W$ state \eqref{Wstate} can be self-tested using only two-party statistics, with three measurements per party. Moreover, the self-testing is robust.
\end{result}

We consider the scenario in which each party (Alice, Bob, and Charlie) performs three dichotomic measurements denoted $Z$, $X$, and $D$, with outcomes denoted as $\pm 1$. Suppose that the observed behavior exhibits the following one- and two-body statistics:
\begin{align}
\label{eq:wCorrelator}
    \braket{Z_m} &=\frac{1}{3},& \braket{X_m}&=0,& \braket{D_m} &= \frac{1}{3\sqrt2},\nonumber\\
    \braket{Z_mZ_n} &= -\frac{1}{3},& \braket{Z_mX_n} &= 0,& \braket{Z_mD_n} &=-\frac{1}{3\sqrt2},\nonumber\\
    \braket{X_mX_n} &=\frac{2}{3}，& \braket{X_mD_n}&=\frac{\sqrt2}{3},& \braket{D_mD_n}& = \frac{1}{6},
\end{align}
where $m,n\in \left\{ A,B,C\right\}$ and $m \neq n$ \footnote{Some marginal statistics (\ref{eq:wCorrelator}) determine the three-party correlators uniquely by sheer inspection. For instance, for all eight $P(a,b,c|Z,Z,Z)$ to be non-negative, it is necessary that $\bra{\psi}Z_AZ_BZ_C\ket{\psi}=-1$. This implies $Z_C\ket{\psi}=-Z_AZ_B \ket{\psi}$, which implies $\bra{\psi}Z_AZ_BX_C\ket{\psi}=0$ and $\bra{\psi}Z_AZ_BD_C\ket{\psi}=-\frac{1}{\sqrt{2}}$; of course, permutations of these hold too. The claim of self-testing implies that actually (\ref{eq:wCorrelator}) determines \textit{all} the three-partite correlators, but we have not been able to find some of the analytical manipulations and self-testing will be proved through the \swap SDP. This is not the first example where even the ideal self-testing is proved with the \swap method: notably, the self-testing of the CGLMP$_3$ inequality in Refs \cite{Yang,Bancal} was also proved only at the level of the SDP.}. These are the statistics that one would obtain for the $W$ state if $Z\equiv \sigma_z$, $X\equiv\sigma_x$ and $D\equiv \frac{1}{\sqrt{2}}(\sigma_z+\sigma_x)$. To investigate the robustness of self-testing induced by these statistics, we shall consider mixing them with white noise, that is by multiplying each term by $(1-\varepsilon)$.

\begin{figure}
  \centerline{
    \begin{tikzpicture}[thick]
    %
    \tikzstyle{operator} = [draw,fill=white,minimum size=1.5em]
    \tikzstyle{phase} = [fill,shape=circle,minimum size=5pt,inner sep=0pt]
    \tikzstyle{surround} = [fill=blue!10,thick,draw=black,rounded corners=2mm]
    \node (Psi) at (-1.5,-2.4) {$\ket{\Psi}_{ABC}$};
    \node at (0,0) (a) {$\ket{0}_a$};
    \node at (0,-0.8) (A) {};
    \node at (0,-1.6) (b) {$\ket{0}_b$};
    \node at (0,-2.4) (B) {};
    \node at (0,-3.2) (c) {$\ket{0}_c$};
    \node at (0,-4) (C) {};
    \draw[-] (Psi) -- (0,-0.8);
    \draw[-] (Psi) -- (0,-2.4);
    \draw[-] (Psi) -- (0,-4);
    %
    \node[operator] (Ha1) at (1,0) {$H$} edge [-] (a);
    \node[operator] (Hb1) at (1,-1.6) {$H$} edge [-] (b);
    \node[operator] (Hc1) at (1,-3.2) {$H$} edge [-] (c);
    %
    \node[phase] (CZa) at (1.8,0) {} edge [-] (Ha1);
    \node[operator] (CZA) at (1.8,-0.8) {$Z$} edge [-] (0,-0.8);
    \draw[-] (CZa) -- (CZA);
    \node[phase] (CZb) at (1.8,-1.6) {} edge [-] (Hb1);
    \node[operator] (CZB) at (1.8,-2.4) {$Z$} edge [-] (0,-2.4);
    \draw[-] (CZb) -- (CZB);
    \node[phase] (CZc) at (1.8,-3.2) {} edge [-] (Hc1);
    \node[operator] (CZC) at (1.8,-4) {$Z$} edge [-] (0,-4);
    \draw[-] (CZc) -- (CZC);
    %
    \node[operator] (Ha2) at (2.6,0) {$H$} edge [-] (CZa);
    \node[operator] (Hb2) at (2.6,-1.6) {$H$} edge [-] (CZb);
    \node[operator] (Hc2) at (2.6,-3.2) {$H$} edge [-] (CZc);
    %
    \node[phase] (CXa) at (3.4,0) {} edge [-] (Ha2);
    \node[operator] (CXA) at (3.4,-0.8) {$X$} edge [-] (CZA);
    \draw[-] (CXa) -- (CXA);
    \node[phase] (CXb) at (3.4,-1.6) {} edge [-] (Hb2);
    \node[operator] (CXB) at (3.4,-2.4) {$X$} edge [-] (CZB);
    \draw[-] (CXb) -- (CXB);
    \node[phase] (CXc) at (3.4,-3.2) {} edge [-] (Hc2);
    \node[operator] (CXC) at (3.4,-4) {$X$} edge [-] (CZC);
    \draw[-] (CXc) -- (CXC);
    %
    \node (enda) at (4.2,0) {} edge [-] (CXa);
    \node (endA) at (4.2,-0.8) {} edge [-] (CXA);
    \node (endb) at (4.2,-1.6) {} edge [-] (CXb);
    \node (endB) at (4.2,-2.4) {} edge [-] (CXB);
    \node (endc) at (4.2,-3.2) {} edge [-] (CXc);
    \node (endC) at (4.2,-4) {} edge [-] (CXC);
    %
    %
    \end{tikzpicture}
  }
  
  \caption{The \swap circuit. The local isometry used to self-test the $W$ state. $H$ is the standard Hadamard gate, $Z$ and $X$ are controlled by the auxiliary qubit. The trusted ancillary qubits are prepared in the state $\ket{0}$.}
  \label{fig:swap}
\end{figure}
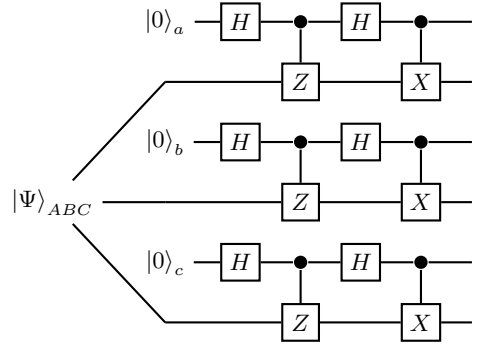

We consider the same isometry as Ref.~\cite{Wu} as shown in Fig.~\ref{fig:swap}.
%
After this isometry, the trusted auxiliary systems will be left in the state
\begin{align}
\rho _{swap}&=\Tr_{ABC}[U\rho_{ABC}\otimes\ket{000}\bra{000}_{A'B'C}U^\dagger]\nonumber\\
&=\sum
C_{ijklst} \ket{i}\bra{j}\otimes\ket{k}\bra{l}\otimes\ket{s}\bra{t}\nonumber
\end{align}
where
\begin{align}
C_{ijklst}=\frac{1}{64}\Tr_{ABC}[M^A_{(j,i)}\otimes M^B_{(l,k)} \otimes M^C_{(t,s)}\rho_{ABC}]\nonumber
\end{align}
and $M^A_{(j,i)}=(I+Z_A)^{j+1}(X_A-X_AZ_A)^{j}(I+Z_A)^{i+1}(X_A-X_AZ_A)^{i}$ for $i,j\in\{0,1\}$, and the expressions for 
$B$ and $C$ are analogous. Finally, we shall be able to express the fidelity
$f=\bra{W}\rho_{swap}\ket{W}$ as a linear function of correlators.

\begin{figure}[tb]
\centering
\includegraphics[scale=0.45]{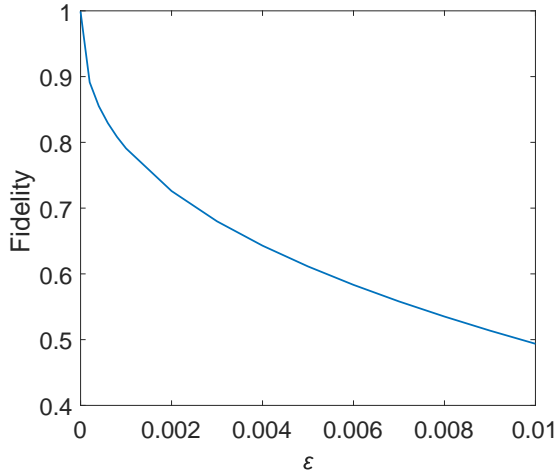}
\caption{Swap bound on the fidelity of the $W_3$ state for different $\varepsilon$. 
$\varepsilon$ represents the deviation of the observed behavior from the ideal values. Notice that although the \swap circuit does not contain the measurements $D_A$, $D_B$ and $D_C$, their appearance in the NPA matrix is crucial for the bound on the fidelity.}
\label{fig:robustness}
\end{figure}

Due to the symmetry present in $\ket{W}$, the measurements as well as the \swap operation, the constraints and the objective function are also symmetric. Hence we can reduce the number of variables in the SDP by solving it in a symmetric space \cite{rosset2018symdpoly,tavakoli2018enabling}. 
%
Let $G= \{ g_i\}_{i=0,\cdots, 5} = \{ (),(ab),(ac),(bc),(acb),(abc)\}$ be the permutation group of three elements. The effect of $g$ on the operators are as expected: for example, $g_5[f(Z_A,X_B,D_C)] = f(X_A,D_B,Z_C)$. The effect of the operator is overloaded to the matrix $\Gamma$ as permutations of rows and columns. So now one can solve the SDP with a symmetrized NPA matrix:
\begin{align*}
    \tilde{\Gamma} = \frac{1}{6}\sum_{i=0}^5 g_i[\Gamma].
\end{align*}
Then we solve the following SDP: 
\begin{align}
\label{eq:wSDP}
\min \quad & f=\bra{W}\rho _{swap}|\ket{W} \nonumber\\
\text{s.t.}\quad  &\tilde{\Gamma} \geq 0, \\
& [\text{equations}~\eqref{eq:wCorrelator}]\times (1-\varepsilon), \nonumber
\end{align}
where $\Gamma$ is a $125\times 125$ NPA matrix of so-called local level one and augmented by necessary terms to express the fidelity. The fidelity is $99.991\%$ when $\varepsilon=0$ and for other $\varepsilon$ up to $0.01$ is shown in Fig.~\ref{fig:robustness}. Unfortunately, the tolerance of noise is so low that this lower bound is hardly relevant to experimental realization.

\begin{figure}[tb]
\centering
\includegraphics[scale=0.45]{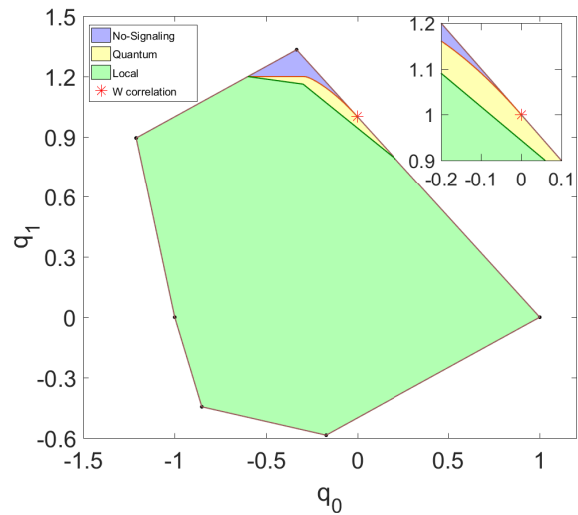}
\caption{The slice of the no-signalling polytope contains $P_W$. 
The point $P_W$ is not exposed in this slice, which means that it is not exposed in the polytope (since an exposed point would be exposed in every slice).}
\label{fig:Wslice}
\end{figure}

From the dual of the SDP \eqref{eq:wSDP} in the ideal case $\varepsilon=0$, 
we can extract a permutationally-invariant Bell inequality $B$ with three measurements and two outputs per party that achieves its maximal value for the observed correlations \eqref{eq:wCorrelator}. The generic form of such an inequality is
\begin{equation}
\label{eq:ineqDual}
    \begin{array}{ll}B&=\alpha S_0+\beta S_1+\gamma S_2
    +\lambda_0T_{00}+\lambda_1T_{11}\\\\
    &+\lambda_2T_{22}+\omega_0T_{01}+\omega_1T_{02}+\omega_2T_{12},
\end{array}
\end{equation}
where $S_i= \langle M_i^A\rangle+\langle M_i^B\rangle+\langle M_i^C\rangle$ 
, $T_{ii}=\langle M_i^AM_i^B\rangle+\langle M_i^BM_i^C\rangle+\langle M_i^CM_i^A\rangle$
and 
$T_{ij}=\langle M_i^AM_j^B\rangle+\langle M_i^BM_j^C\rangle+\langle M_i^CM_j^A\rangle
+\langle M_j^AM_i^B\rangle+\langle M_j^BM_i^C\rangle+\langle M_j^CM_i^A\rangle$ for $i,j\in\{0,1,2\}$ and $i\neq j$. By inspection, the dual yields $\alpha\approx-\lambda_0$ and much smaller values for all the other coefficients \footnote{The outcome of our program is 
\ban
    &(\alpha,\beta,\gamma,\lambda_0,\lambda_1,\lambda_2,\omega_0,\omega_1,\omega_2)=\\
    &(3.7701225, -0.00000713,0.00083364, \\
    &-3.76273900, 0.00004938,0.00060403,\\
    &-0.00020047, -0.00227982, 0.00027316)\times10^4
\ean}. The resulting guess $B\approx S_0 - T_{00}$ defines a positivity facet, hinting that the correlation~\eqref{eq:wCorrelator} is a \textit{non-exposed} extremal point of the quantum set~\cite{goh2018geometry}. We strengthen the evidence in favor of this conjecture by plotting (Fig.~\ref{fig:Wslice}) the quantum set on a slice of the no-signaling polytope. We take the slice defined by $P(q_0,q_1)=q_0P_{\text{local}}+q_1P_W+(1-q_0-q_1)P_{\text{noise}}$,
where $P_W$ are the marginal statistics~\eqref{eq:wCorrelator} 
and $P_{\text{noise}}$ is the maximally mixed behavior, and $P_{\text{local}}$ is the one obtained by taking the local deterministic point $Z_A=-X_A=D_A=-Z_B=X_B=D_B=-Z_C=X_C=-D_C=-1$ and applying all the six permutations of the parties.
This plot provides graphical evidence that the self-testing of that behavior cannot be associated to the maximal violation of a single inequality. 
\footnotetext{test.}

\begin{result}
For all real $\lambda \in (0,1]$, the three-qubit state \begin{equation}
\label{eq:generalW}
|\psi_\lambda\rangle=\frac{1}{\sqrt{2+\lambda^2}}(|100\rangle+|010\rangle+\lambda|001\rangle)
\end{equation}
can be self-tested with only two-body correlators with three measurements per party. 
\end{result}

\begin{figure}[tb]
\centering
\includegraphics[scale=0.45]{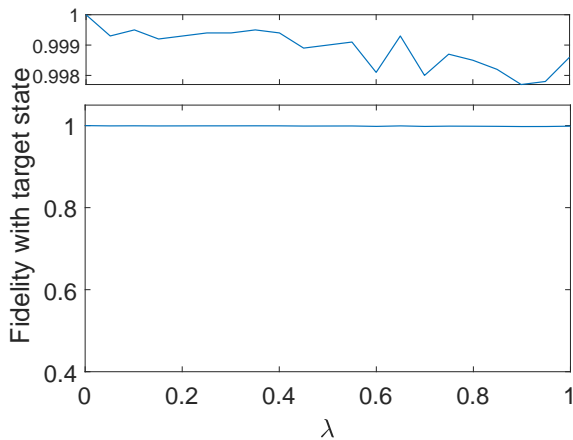}
\caption{Blue line represents the lower bound on the fidelity obtained 
with varying $\lambda$ from zero to 1. When $\lambda=0$, 
the state becomes a product of a Bell state and a qubit, so the fidelity under the marginal is higher than other points.}
\label{fig:lambda}
\end{figure}

Self-testing of these states with three-body correlators was proved in Ref.~\cite{Wu} using two measurements per party. To self-test them with only one- and two-body marginals, we consider the statistics associated to the measurements $(Z,X,D)$ as above; the explicit expressions are written in the Appendix A. As before, the fidelity function is written as a linear combination of observables and SDP variables. Notice that, since the target state depends on $\lambda$, the fidelity is also a function of $\lambda$. Then we minimize the fidelity for various $\lambda \in (0,1]$ using a moment matrix $\Gamma$ of size $83\times83$. For all $\lambda$, the SDP returns $f>99.8\%$ (Fig.~\ref{fig:lambda}): we believe that the deviation from 1 is due to the limitation of the SDP relaxation.

\begin{result}
For three parties with two dichotomic measurements, Ref.~\cite{Sainz} proved that only one non-trivial translationally invariant Bell inequality can be built on one- and two-party statistics. We prove that the maximal violation of that inequality self-tests a three-qubit state. Moreover, the self-testing is robust.
\end{result}

The three-partite Bell inequality proposed under study reads $B\leq 9$ with
\begin{align}
\label{eq:belenIneq}
B = -S_0-3S_1- T_{00}+3 T_{11}+T_{01}+2T_{10},
\end{align}
where
$T_{ij}=\langle M_i^AM_j^B\rangle+\langle M_i^BM_j^C\rangle+\langle M_i^CM_j^A\rangle$, so that the inequality is translationally but not permutationally invariant. Since each party has only two dichotomic measurements, the maximal quantum violation can be achieved with projective measurements on qubits~\cite{masanes2006asymptotic}, which of course does not mean \textit{a priori} that it could not be achieved also by other resources: this is what we set out to prove. Writing the qubit measurements as
$$M_j^{(i)}=\cos \theta^{(i)}_j \sigma_z+\sin \theta^{(i)}_j\sigma_x$$
where $j\in\{0,1\}$, $i\in\{A,B,C\}$ and  $ \theta^{(i)}_j\in [-\frac{\pi}{2}, \frac{\pi}{2}]$, for $\theta_0^i\approx -1.1946$ and $\theta_1^i\approx 0.0957$ one obtains the maximal violation $B \approx 10.02$. We are going to prove that this self-tests the corresponding eigenvector
\begin{align}
\label{eq:belenState}
|\psi\rangle\approx &-0.08(|000\rangle+|111\rangle) \nonumber \\
& -0.5628(|001\rangle+|010\rangle+|100\rangle)\nonumber \\
& +0.1108(|011\rangle+|110\rangle+|101\rangle).
\end{align}

Let us first look at the ideal quantum realization to design our \swap circuit. We can rotate the local bases so that $M_1=\sigma_x$ for Alice, Bob, and Charlie. This sets 
$M_0=\sin(\theta_0+\theta_1)\sigma_z+\cos(\theta_0+\theta_1)\sigma_x$.

In order to construct the \swap circuit, we'd rather need $\sigma_z$, which in the ideal case is 
$[M_0-\cos (\theta_0+\theta_1)M_1]/\sin(\theta_0+\theta_1)$.
However, written with the unknown measurement operators, this expression may not define a unitary operator. A method to circumvent this obstacle has been presented in previous works~\cite{Yang1,Bancal,Wu}: one defines a third dichotomic operator $M_2$ such that \begin{equation}
\label{eq:localizing}
M_2\frac{M_0-\cos (\theta_0+\theta_1)M_1}{\sin(\theta_0+\theta_1)}\geq 0.
\end{equation} Since this equation is not a SDP constraint, one relaxes it to the positivity of a ``localizing matrix''.  

\begin{figure}[tb]
\centering
\includegraphics[scale=0.4]{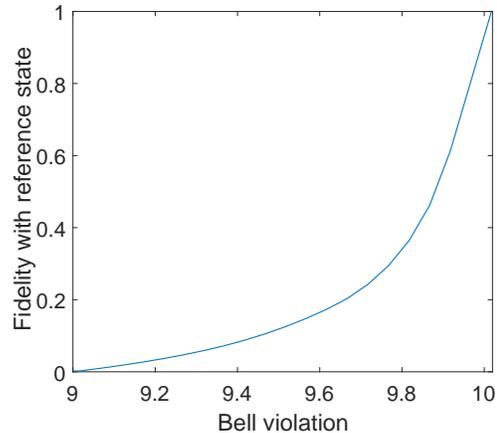}
\caption{Minimal fidelity of the state swapped out of the operators defined above. The blue line represents the lower bound on the fidelity obtained from SDP hierarchy on level-1 with size $88\times 88$.}
\label{fig:belen}
\end{figure}

We ran the SDP, with matrix size $88\times 88$ and augmented by three localizing matrices (one per party), minimizing the fidelity with the maximal violation states, for different magnitude of violation of the inequality~\eqref{eq:belenIneq}. The result is summarized in Fig.~\ref{fig:belen}.

\begin{result}
For $n=4$ parties, the state 
 \begin{equation}\label{eq:W4}
   \ket{W_4}=\frac{1}{\sqrt{4}}(\ket{0001}+\ket{0010}+\ket{0100}+\ket{1000})
 \end{equation}
can be self-tested with three-body correlations and three measurements per party. Moreover, the self-testing is robust.
\end{result}
Like the $W_3$ state, we still use measurements $\left\{Z,X,D\right\}$ each party to construct the \swap circuit. The correlators for the ideal case are given in Appendix B. Then the entries of $\rho_{swap}$ are expressed as linear combinations of correlation terms from the set $c=\{\mathbb{I},\Tr(\rho Z_A),\Tr(\rho Z_AX_B),\cdots,\Tr(\rho Z_AX_AZ_B Z_C X_CZ_D)\}$. We ran the SDP program with the NPA moment matrix with size $167\times 167$. The fidelity $f>99.998\%$ without noise and the robustness is given by the all correlations for ideal values multiplied by $(1-\varepsilon)$, where $\varepsilon$ represents the deviation of the observed behavior from the ideal values (Fig.6).

If we were to use only the two-body correlators for the same measurements, the fidelity would drop below $30\%$.

In view of these observations, we conjecture that the state $\ket{W_n}$ can be self-tested using these three measurements if $(n-1)$-body correlators are given. Whether the same state can be self-tested from fewer-body correlators, using different (and possibly more) measurements, remains an open question.

\begin{figure}[tb]
\centering
\includegraphics[scale=0.4]{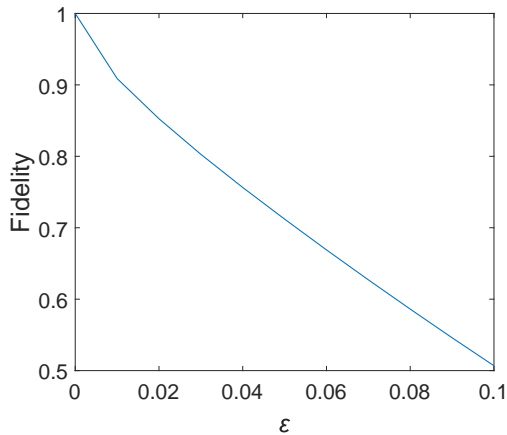}
\caption{Swap bound on the fidelity of the $W_4$ state for different $\varepsilon$.} 

\label{fig:W4robustness}
\end{figure}

\section{Conclusion}

In a multipartite entangled state, a lot of information may be encoded in the partial state---at times, all of it. This observation was known in the context of entanglement theory for characterized degrees of freedom. We have shown that it carries over to the device-independent framework of uncharacterized devices.

The examples we presented all deal with the multipartite scenario and end up self-testing states of three or four qubits. Our work calls for generalization both in local dimensionality and in number of parties. In the tomographic scenario, it it known that $N$-qubit $W$ states can be determined by their bipartite marginals~\cite{parashar2009n} and multipartite $W$-type state is determined by its single-particle reduced density matrices among all $W$-type states~\cite{yu2013multipartite}. A question that may be asked is: up to which number of parties $N$ can one find states that can be self-tested with only marginal information on two-party correlators? This would be important in the context of many-body physics, where the quantities that are routinely measured don't go beyond functions of two-body correlations.

\section*{Acknowledgments}
We would like to thank Jean-Daniel Bancal, Matteo Fadel, Yeong-Cherng Liang, Stefano Pironio, Ana Bel\'en Sainz, Yukun Wang and Xingyao Wu for useful discussions; and Denis Rosset for sharing notes on SDP with group symmetries.

This research is supported by the Singapore Ministry of Education Academic Research Fund Tier 3 (Grant No. MOE2012-T3-1-009), by the National Research Fund and the Ministry of Education, Singapore, under the Research Centres of Excellence programme, and by the John Templeton Foundation Grant No.60607 ``Many-box locality as a physical principle''. This work is also funded by National Nature Science Foundation of China (Grants No.61671082, No.61672110, and No.61572081) and by China Scholarship Council.

\bibliography{references}

\section*{Appendix A}
This appendix provides the details of two-body correlations of 
the state \eqref{eq:generalW} in Result 2 with three dichotomic measurements for each party for $\lambda\in(0,1]$. 

The state being symmetric in $A$ and $B$, it is convenient to list the correlators in three sets:

\textbf{Set 1.} For the parties $A$ and $B$, with $m,n=\{A,B\}$ and $m\neq n$: 
\ban
\braket{Z_m}&=&\frac{\lambda^2}{\lambda^2+2}\\
\braket{Z_mZ_n}&=&\frac{\lambda^2-2}{\lambda^2+2}\\
\braket{Z_mD_n}&=&\frac{\lambda^2-2}{\sqrt{2}(\lambda^2+2)}\\
\braket{X_mX_n}&=&\frac{2}{\lambda^2+2}\\
\braket{D_mD_n}&=&\frac{\lambda^2}{2(\lambda^2+2)}\,.
\ean

\textbf{Set 2.} For either $A$ or $B$ together with $C$, i.e. with $m\in\{A,B\}$: 
\ban
\braket{Z_C}&=&\frac{2-\lambda^2}{\lambda^2+2}\\
\braket{Z_mZ_C}&=&\frac{-\lambda^2}{\lambda^2+2}\\
\braket{Z_mD_C}&=&\frac{-\sqrt{2}\lambda^2}{2(\lambda^2+2)}\\
\braket{X_mX_C}&=&\frac{2\lambda}{\lambda^2+2}\\
\braket{X_mD_C}&=&\frac{\sqrt{2}\lambda}{\lambda^2+2}\\
\braket{D_mZ_C}&=&-\frac{\sqrt{2}\lambda^2}{2(\lambda^2+2)}\,.
\ean

\textbf{Set 3.} For any two parties, i.e. $m,n\in\{A,B,C\}$ and $m\neq n$:
\ban
\braket{X_m}&=&0\\
\braket{D_m}&=&\frac{\braket{Z_m}}{\sqrt{2}}\\
\braket{D_mX_n}&=&\frac{\braket{X_mX_n}}{\sqrt{2}},\\
\braket{Z_mX_n}&=&0\\
\braket{Z_mD_n}&=&\frac{\braket{Z_mZ_n}}{\sqrt{2}}\\
\braket{D_mD_n}&=&\frac{\braket{Z_mZ_n}+\braket{X_mX_n}}{2}\,.
\ean

\section*{Appendix B}
The state \eqref{eq:W4} in Result 4 is symmetric for four parties. So the correlations can be divided into three sets:\\

\textbf{Set 1.} For any one party, i.e. $m\in\{A,B,C,D\}$:
\begin{equation*}
\langle Z_m\rangle= \frac{1}{2};~~\langle X_m\rangle=0;~~\langle D_m\rangle=\frac{1}{2\sqrt{2}};
\end{equation*}

\textbf{Set 2.} For any two parties, i.e. $m,n\in\{A,B,C,D\}$ and $m\neq n$:
\begin{equation*}\begin{array}{ll}
&\langle Z_mZ_n\rangle=0;~~\langle X_mX_n\rangle=\frac{1}{2};~~\langle D_mD_n\rangle=\frac{1}{4}; \\\\
&\langle Z_mX_n\rangle=0;~~\langle X_mD_n\rangle=\frac{1}{2\sqrt{2}};~~\langle D_mZ_n\rangle=0;
\end{array}
\end{equation*}

\textbf{Set 3.} For any three parties, i.e. $m,n,k\in\{A,B,C,D\}$ and $m\neq n\neq k$:
\begin{equation*}\begin{array}{ll}
&\langle Z_mZ_nZ_k\rangle=-\frac{1}{2};~~\langle X_mX_nX_k\rangle=0;~~\langle D_mD_nD_k\rangle=\frac{1}{2\sqrt{2}}; \\\\
&\langle Z_mZ_nX_k\rangle=0;~~\langle Z_mZ_nD_k\rangle=-\frac{1}{2\sqrt{2}};~~\langle X_mX_nZ_k\rangle=\frac{1}{2}; \\\\
&\langle X_mX_nD_k\rangle=\frac{1}{2\sqrt{2}};~~\langle D_mD_nZ_k\rangle=0;~~\langle D_mD_nX_k\rangle=\frac{1}{2};\\\\ 
&\langle Z_mX_nD_k\rangle=\frac{1}{2\sqrt{2}}.
\end{array}\end{equation*}

\end{document}